\documentclass[openany]{nature}
\usepackage[T1]{fontenc}
\usepackage[left]{lineno}
\usepackage{amsthm,amsmath,amssymb}
\usepackage{mathrsfs}
\usepackage{overpic}
\usepackage[T1]{fontenc}
\usepackage{graphicx}
\usepackage{pdflscape}
\usepackage{hyperref}
\usepackage{threeparttable}
\usepackage{xcolor}
\usepackage{ulem}
\usepackage{cancel}
\usepackage[figuresleft]{rotating}
\usepackage{lineno}
\usepackage{caption}
\usepackage{subfigure}
\usepackage{tablefootnote}
\usepackage{amsmath, amssymb}

\makeatletter

\let\saved@includegraphics\includegraphics
\AtBeginDocument{\let\includegraphics\saved@includegraphics}
\renewenvironment*{figure}{\@float{figure}}{\end@float}
\makeatother

\newcommand{\apj}{Astrophys. J.}
\newcommand{\apjl}{Astrophys. J.}
\newcommand{\apjs}{Astrophys. J.}
\newcommand{\aap}{Astron. Astrophys.}
\newcommand{\mnras}{Mon. Not. R. Astron. Soc.}
\newcommand{\nat}{Nature}
\newcommand{\araa}{Ann. Rev. Astron. Astrophys.}

\def\be{\begin{eqnarray}}
\def\ee{\end{eqnarray}}

\makeatletter

\def\@fnsymbol#1{\ensuremath{\ifcase#1\or \dagger\or \ddagger\or
 \mathsection\or \mathparagraph\or \|\or **\or \dagger\dagger
 \or \ddagger\ddagger \else\@ctrerr\fi}}
\makeatother

\usepackage{hyperref}
\usepackage{graphicx}
\usepackage{longtable}
\usepackage{supertabular}
\usepackage{float} 
\usepackage{authblk}

\title{A Peculiarly Short-duration Gamma-Ray Burst from Massive Star Core Collapse}

\author{B.-B. Zhang$^{1,2,3,\ast}$\thanks{E-mail: bbzhang@nju.edu.cn}, Z.-K. Liu$^{1,2,\ast}$, Z.-K. Peng$^{1,2,\ast}$, Y. Li$^{4}$, H.-J. L\"u$^{5}$, J. Yang$^{1,2}$, Y.-S. Yang$^{1,2}$, Y.-H. Yang$^{1,2}$, Y.-Z. Meng$^{1,2}$, J.-H. Zou$^{1,2,6}$, H.-Y. Ye$^{1}$, X.-G. Wang$^{5}$, J.-R. Mao$^{7}$, X.-H. Zhao$^{7}$, J.-M. Bai$^{7}$, A. J. Castro-Tirado$^{8,9}$, Y.-D. Hu$^{8,10}$, Z.-G. Dai$^{11,1,2}$, E.-W. Liang$^{5}$, B. Zhang$^{3}$\thanks{E-mail: zhang@physics.unlv.edu}}

\begin{document}
\maketitle
\begin{affiliations}
 \item School of Astronomy and Space Science, Nanjing University, Nanjing 210093, China
 \item Key Laboratory of Modern Astronomy and Astrophysics (Nanjing University), Ministry of Education, China
 \item Department of Physics and Astronomy, University of Nevada, Las Vegas, NV 89154, USA
\item Kavli institute for astronomy and astrophysics, Peking University, Beijing. 100871, P.R.China
 \item Guangxi Key Laboratory for Relativistic Astrophysics, School of Physical Science and Technology, Guangxi University, Nanning 530004 China;
 \item College of Physics, Hebei Normal University, Shijiazhuang 050024, China
\item Yunnan Observatories, Chinese Academy of Sciences, 650216, Kunming, China
 \item Instituto de Astrof\'isica de Andaluc\'ia (IAA-CSIC), Glorieta de la Astronom\'ia s/n, E-18008, Granada, Spain
 \item Departamento de Ingenier\'ia de Sistemas y Autom\'atica, Escuela de Ingenier\'ias, Universidad de M\'alaga, Dr. Pedro Ortiz Ramos, 29071 M\'alaga, Spain
 \item Universidad de Granada, Facultad de Ciencias Campus Fuentenueva s/n E-18071 Granada, Spain
\item Department of astronomy, School of Physical Science, University of Science and Technology of China, Hefei 230026, Anhui, China
 \end{affiliations}
\noindent{$^\ast$Co-First Authors. These authors contributed equally:Bin-Bin Zhang, Zi-Ke Liu and Zong-Kai Peng }
\bigskip

\begin{abstract} 
\bf Gamma-ray bursts (GRBs) have been phenomenologically classified into long and short populations based on the observed bimodal distribution of duration\cite{1993ApJ...413L.101K}. Multi-wavelength and multi-messenger observations in recent years have revealed that in general long GRBs originate from massive star core collapse events\cite{Woosley2006}, whereas short GRBs originate from binary neutron star mergers\cite{Abbott2017ApJ...848L..12A}. It has been known that the duration criterion is sometimes unreliable, and multi-wavelength criteria are needed to identify the physical origin of a particular GRB\cite{2009ApJ...703.1696Z1}. Some apparently long GRBs have been suggested to have a neutron star merger origin\cite{2006Gehrels}, whereas some apparently short GRBs have been attributed to genuinely long GRBs\cite{levesque10} whose short, bright emission is slightly above the detector's sensitivity threshold. Here we report the comprehensive analysis of the multi-wavelength data of a bright short GRB 200826A. 
Characterized by a sharp pulse, this burst shows a duration of 1 second and no evidence of an underlying longer-duration event. 
Its other observational properties such as its spectral behaviors, total energy, and host galaxy offset, are, however, inconsistent with those of other short GRBs believed to originate from binary neutron star mergers. Rather, these properties resemble those of long GRBs. This burst confirms the existence of short duration GRBs with stellar core-collapse origin\cite{2009ApJ...703.1696Z1}, and presents some challenges to the existing models. 
\end{abstract}

\begin{figure}
\hspace{-1.5cm}
\includegraphics[width=8.0in]{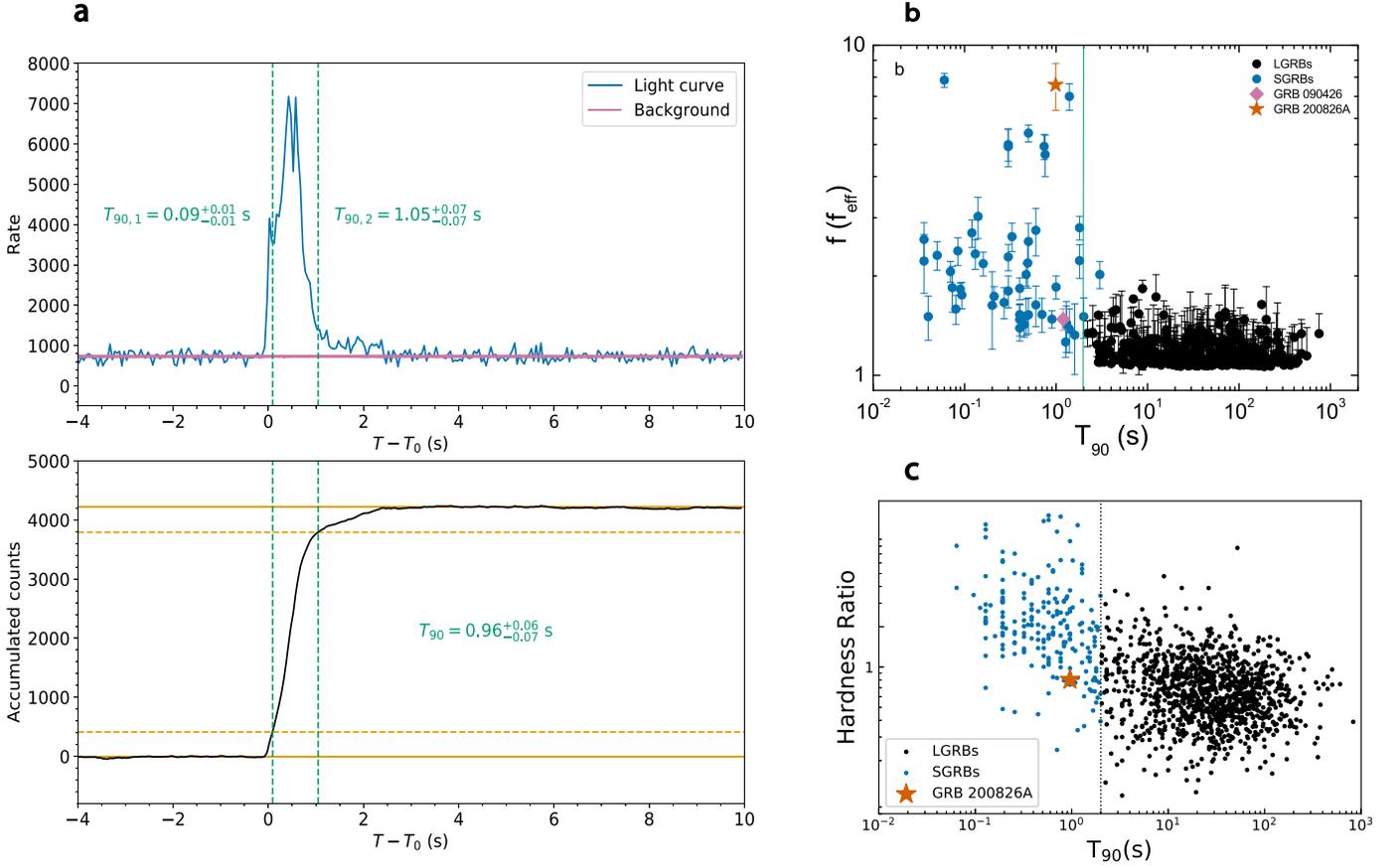}

\caption{{\bf Temporal properties of GRB 200826A. a,} $T_{\rm 90}$ calculation : The upper panel shows the light curve and background flux level with blue and purple solid lines, respectively. The accumulated counts as a function of time are shown with the black curve in the lower panel. The orange horizontal dashed (solid) lines are drawn at 5$\%$(0$\%$) and 95$\%$ (100$\%$) of the total accumulated counts. The $T_{90}$ interval is marked by the green vertical dashed lines. {\bf b,} $T_{90}-f(f_{\rm eff})$ plot. The $f$ parameter measures the ratio between peak flux average background flux of a GRB (Methods). $f_{\rm eff}$ is the effective $f$ parameter when a long GRB becomes a disguised short GRB by arbitrarily lowering its flux level. GRB 200826A is highlighted by the orange solid star. Blue circles correspond to $f$ parameter values of short GRBs, black circles mark $f_{\rm eff}$ values of long GRBs and green vertical line is the division line at $ 2~\rm s$. {\bf c,} GRB 200826A on the $T_{\rm 90}$-HR diagram : Blue and black points represent the short and long GRBs, respectively; orange solid star marks GRB 200826A; and the black dashed line separate long and short GRBs at $2~\rm s$. All error bars represent 1-$\sigma$ uncertainties.
}
\label{fig1}
\end{figure}

GRB 200826A triggered the {$Fermi$} Gamma-ray Burst Monitor (GBM)\cite{2009ApJ...702..791M} at 04:29:52 Universal Time on 26 August 2020.
\cite{GCN28287}.
Follow-up observations detected the optical counterpart
\cite{GCN28295}
and identified its host galaxy and redshift at $z=0.7481$.
(ref. \cite{GCN28319}). 
The 10-800 keV-band light curve, as detected by GBM, is shown in Figure \ref{fig1}a. The duration in the same energy range of the burst is measured
as $T_{90} \sim0.96_{-0.07}^{+0.06}$ s (see Table 1 for a collection of all the measured parameters of GRB 200826A, and see the details of data analysis in Methods), which is consistent with the value of $T_{90}=1.14\pm0.13$ in 50-300 keV energy range\cite{ahumada2020}. The lightcurve pulse profile is quite sharp. There is no signal above the background both before and after the burst, hinting that the short duration is genuine. We further measure the ``amplitude parameter'' of this burst\cite{lv14}, which is defined as the ratio between the peak flux and the average background flux of the GRB lightcurve (see Methods). We obtain $f=7.58 \pm 1.23$ for GRB 200826A. To make a long GRB a ``tip-of-iceberg'' short GRB, the effective $f$ value is typically $<1.5$ (ref. \cite{lv14}). The disguised short GRB 090426 with a massive star origin\cite{levesque10} had $f=1.48 \pm 0.11$, which is consistent with having underlying long-duration emission not observed above the detection level, although its origin remains an open question due to its low hydrogen column density and weak Ly-$\alpha$ lines\cite{2009A&A...507L..45A}. The enormous $f$ value of GRB 200826A, therefore, suggests that it is a genuinely short GRB and cannot be the tip-of-iceberg of a long GRB (Figure.\ref{fig1}b). 

Figure \ref{fig1}c shows the standard long/short classification 
diagram in the duration - hardness ratio domain, where the hardness ratio is defined as observed counts ratio between the 50-300 keV band and the 10-50 keV band within the $T_{90}$ duration (see Methods). One can see that GRB 200826A falls into the distribution of short GRBs, even though near the softer end of the hardness ratio distribution (with a value 0.803). We perform the spectral analysis of the GRB (see Fermi data analysis in Methods and \textcolor{black}{Extended Data Figure 1}). \textcolor{black}{Its time-averaged photon index is $\alpha= -0.68\pm0.05$ , spectral peak energy is $E_{\rm p} \simeq 120.29_{-3.67}^{+3.93}$ keV and they show the same evolutionary pattern as light curves (Extended Data Figure 2 and Extended Data Figure 3).} These are not too different from other short GRBs. Without the redshift and host galaxy information, this burst would be classified as a member in the short GRB population.

\begin{figure}
 \vspace{-2.3cm}
\begin{center}
\includegraphics[scale=0.7]{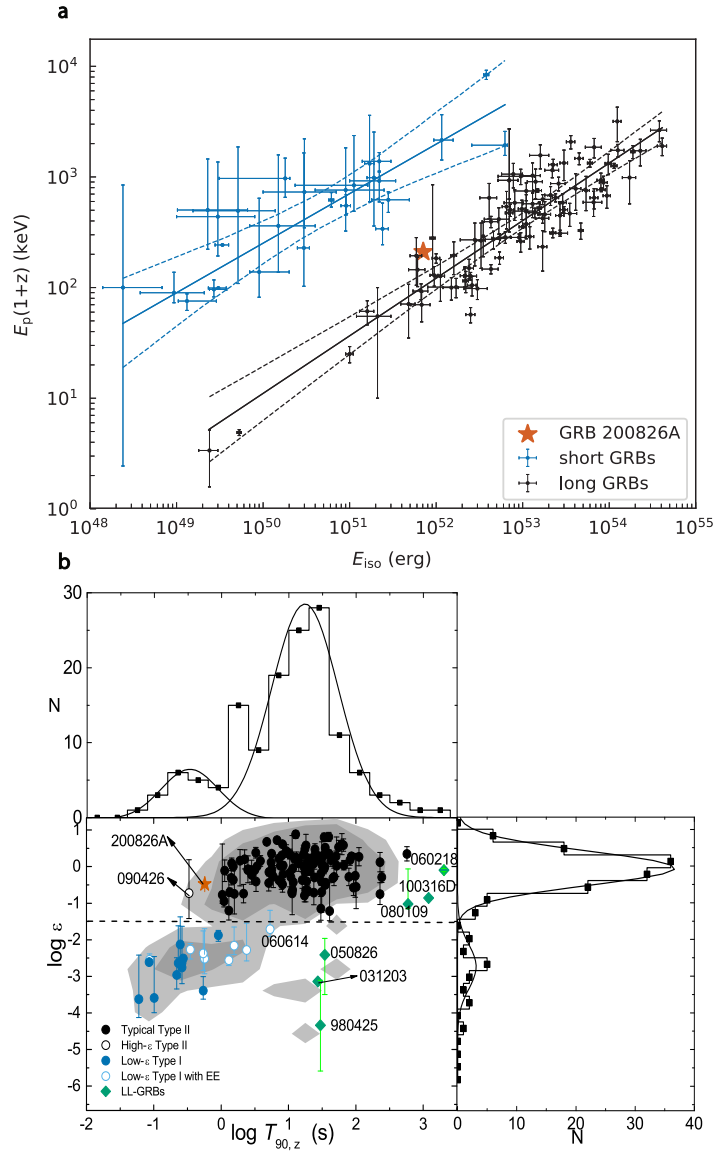}
\end{center}

\caption{{\bf GRB 200826A in energy-related correlations. a,} $E_{\rm p,z} - E_{\rm \gamma,iso}$ correlation diagram for short and long GRBs. Dashed borderlines show the 3$\rm \sigma$ regions for each correlation. {\bf b,} One and two dimensional distributions of GRB samples in the $\epsilon - T_{90}$ space. {\bf$\epsilon$} is defined as {\bf$\epsilon$} = $E_{\rm \gamma,iso,52}/E_{\rm p,z,2}^{5/3}$, where $E_{\rm \gamma,iso,52}=E_{\rm \gamma,iso}/10^{52}~\rm erg$ and $E_{\rm p,z,2}=E_{\rm p,z}/100~\rm keV$ ( See Methods. Blue and black solid circles represent the Type I and Type II GRB candidates, respectively. The special case of the short-duration GRB 090426 with a high-$\rm \epsilon$ is marked by an open black circle. Green diamonds denote the nearby low-luminosity long GRBs. Probability contours of Type I and II GRB clusters are also displayed in the grey regions (see Methods). The dashed horizontal line is the $\epsilon$ = 0.03 division line. The Gaussian fits to the distributions of $\epsilon$ and $T_{90}$ of the two populations are plotted as black curves. \textcolor{
black}{Contours with 2-$\sigma$ and 3-$\sigma$ level are marked in grey}. One can see that GRB 200826A (marked as a \textcolor{black}{orange} star) comfortably falls in the Type II region in both plots. All error bars represent 1-$\sigma$ uncertainties.
}
\label{fig2}
\end{figure}

The conclusion changes when the redshift information is considered. With the measured redshift, we derive an isotropic energy of $E_{\rm \gamma,iso} \simeq 7.09\pm 0.28\times10^{51}$ erg based on the measured fluence of $4.85\pm 0.19\times 10^{-6} {\rm erg\hspace{3pt}cm^{-2}}$. This energy value is relatively large compared with other short GRBs but is typical in the long GRB population. Such a discrepancy is more obvious when GRB 200826A is plotted in the $E_{\rm \gamma,iso} - E_{\rm p,z}$ relation for both long and short GRBs (where $E_{\rm p,z}=E_{\rm p} (1+z)$ is the spectral peak energy in the burst rest frame). Observations showed that there is an empirical relation between the two, the so-called Amati-relation\cite{Amati2002} (see Methods). Long and short GRBs are known to follow the same trend but form distinct tracks\cite{2009ApJ...703.1696Z1}. We find that GRB 200826A squarely lies on the long GRB track rather than on the short GRB track (Figure \ref{fig2}a). 

The peculiarity of GRB 200826A is also reflected in its value of the 
$\epsilon \equiv E_{\rm \gamma,iso,52}/E_{\rm p,z,2}^{5/3}$ parameter\cite{2010ApJ...725.1965L}, where $E_{\rm \gamma,iso,52}$ is the isotropic gamma-ray energy in units of $10^{52}$ erg, and $E_{\rm p,z,2}$ is $E_{p,z}$ in units of 100 keV. Following the convention of ref. \cite{Zhang06Nature}, hereafter we define Type I and Type II GRBs as those with a compact star merger origin and a massive star core collapse origin, respectively. It has been shown that GRBs can be more distinctly classified into their respective physical categories (Type I and Type II) in the $\epsilon - T_{90}$ space, with a separation line of $\epsilon \sim 0.03$ (ref. \cite{2010ApJ...725.1965L}). We obtain $\epsilon = 0.33$ for GRB 200826A and find that it statistically falls into the Type II rather than the Type I category (Figure \ref{fig2}b, see Methods).

Long GRBs usually show ``spectral lags'', with the softer emission peaking later than the harder emission\cite{norris96}. Short GRBs, on the other hand, typically show negligible lags\cite{yi06}. The (10-20) keV to (250-300) keV spectral lag of GRB 200826A is 0.157$\pm 0.051$ s (\textcolor{black}{See Methods and Extended Data Figure 4}). This is at odds for short GRBs but is typical for long GRBs.

\begin{figure}
\vspace{-3.cm}

 \hspace{-1.4cm}
\includegraphics[scale=1.2]{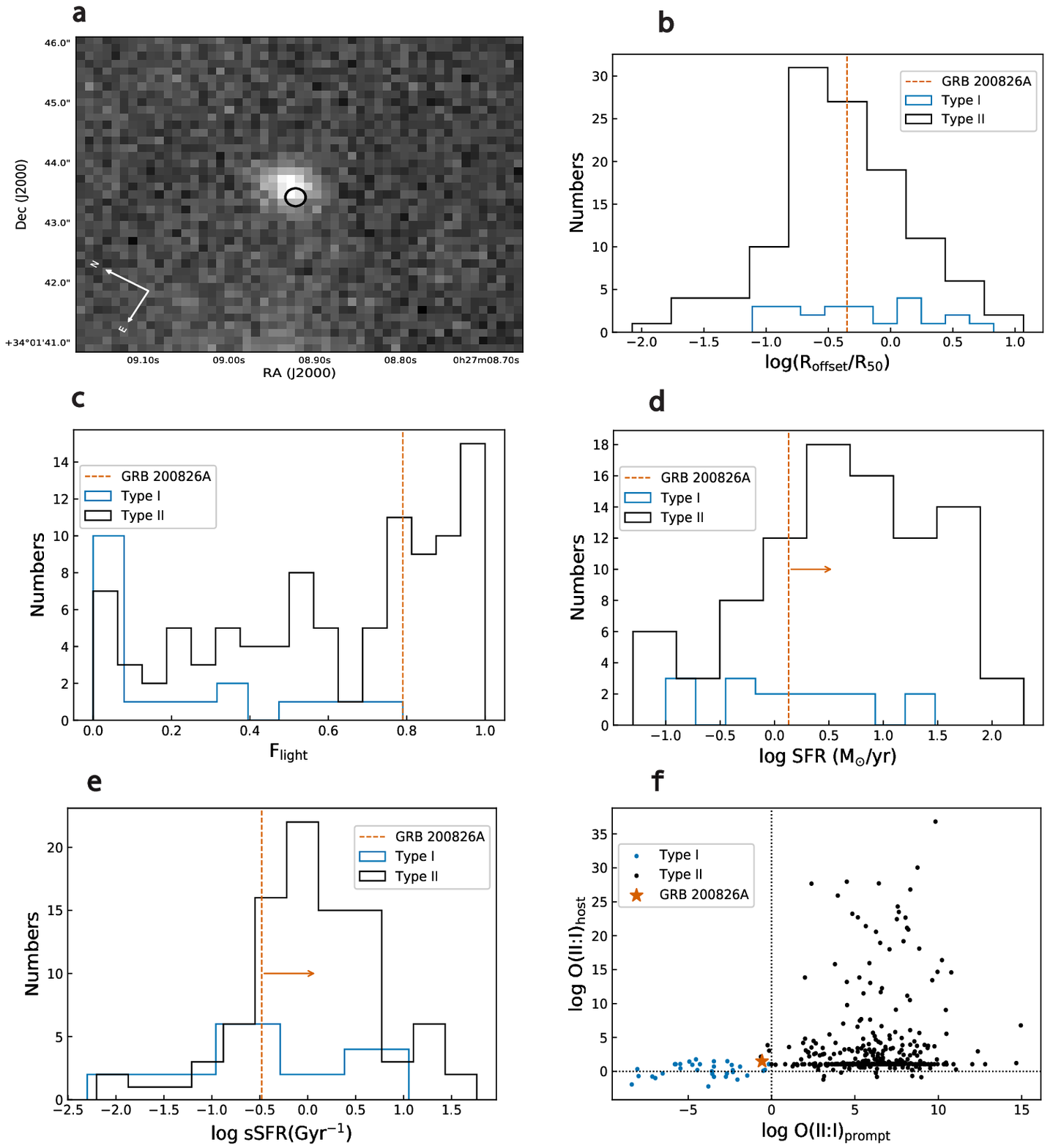}

\caption{{\bf Host properties of GRB 200826A. a,} The Gran Telescopio CANARIAS (GTC) Image of GRB 200826A's host galaxy. {\bf b, }Normalized offset ($r=R_{\rm off}/R_{50}$) distributions of our samples. {\bf c, } Distributions of cumulative light fraction $F_{\rm light}$. {\bf d,} Star formation rate (SFR) of the host galaxies. {\bf e,} specific star formation rate (sSFR; the ratio between SFR
and stellar mass of host galaxy $M_*$) of the host galaxies. {\bf f,} Posterior odds O (II:I) of prompt emission and host properties. GRB 200826A is marked with a \textcolor{black}{orange} star. In {\bf b, c, d, and e}, the black and blue histograms are for Type I and Type II GRBs, respectively. The vertical dashed \textcolor{black}{orange} lines represent the location of GRB 200826A.
}
\label{fig3}
\end{figure}

Binary neutron star mergers are the leading engines for short-duration GRBs. These mergers usually happen after a long delay since star formation so that the SFR and sSFR of their host galaxies are usually not high. The host galaxy and the position (see Methods) of the GRB inside the galaxy often gives the clue regarding the progenitor of a GRB\cite{fruchter06,berger14,liye2020}. We performed a follow-up observation of GRB 200826A using Gran Telescopio Canarias (GTC) and identified the host galaxy (Figure 3a) which was found to lie at redshift $0.7481\pm 0.0003$ (ref. \cite{ahumada2020}). We measured its half-light radius as 0.77", corresponding to 5.7 kpc, in our GTC $r$-band image. The optical afterglow of GRB 200826A has a 0.35" offset from the center of the host galaxy, which corresponds to 2.56 kpc. The normalized offset is $r=R_{\rm off}/R_{50}=0.45$. We estimate the cumulative light fraction $F_{\rm light}=0.79$. We also calculated the star formation rate (SFR; see Methods) to be {$>1.44\ M_{\odot} {\rm yr^{-1}}$} and specific star formation rate (sSFR) to be {$>0.35\ {\rm Gyr^{-1}}$}. Our results are consistent with those in ref.\cite{ahumada2020}. Due to kicks at the births of neutron stars from SNe explosions, these mergers also usually occur at a site far away from star formation regions, often in the outskirts or even outside their host galaxies\cite{gehrels05,berger14}. In Figure \ref{fig3}b-e, we place GRB 200826A in the $r$, $F_{\rm light}$, SFR, and sSFR distributions for long (Type II) and short (Type I) GRBs\cite{liye2016}. Statistically speaking, one can see that the host galaxy properties of GRB 200826A are more consistent with belonging to typical Type II GRBs rather than Type I.

To make our conclusion more quantitative, we apply the method of ref.\cite{liye2020} that classifies GRBs into physical categories using multi-wavelength observational data (see Methods). The logarithmic odds $\log O {\rm( II:I)_{host}}$ using host galaxy information only is 2.5, indeed consistent with a Type II origin. When the prompt emission information is considered, if one ignores the $f$ information, the overall indicator $\mathcal{O}$, which is defined as the logarithmic probability ratio between Type II and Type I GRBs plus 0.7 and assigns Type II/I GRBs as positive/negative values, is 1.5. This places GRB 200826A into the Type II (massive star core-collapse) GRB category (Figure \ref{fig3}f). When the very large $f$ value ($f=7.58\pm1.23$, a character of Type I GRBs) is added as one of the criteria, the method of ref.\cite{liye2020} gives $\mathcal{O} =-3.1$, still placing GRB 200826A in the Type I category. 
This again indicates the uniqueness of GRB 200826A, which points towards a genuinely short burst with other properties consistent with being a massive-star-core-collapse event, so that the GRB classification scheme requires further refinements.

Finally, a Type II origin of GRB 200826 is further supported by the follow-up observations by Gemini-North 8-meter telescope\cite{ahumada2020}\textcolor{black}{(see Extended Data Figure 5)}, which revealed a possible optical bump at $25.45\pm 0.15$ mag in $i$-band. This bump could be consistent with a supernova, a signature of a massive star core collapse event.

The fact that GRB 200826A is genuinely short but its many properties are inconsistent with being of a neutron star merger origin raises a great challenge in identifying its progenitor star. For an accretion-powered GRB engine, the shortest timescale to power a relativistic jet is defined by the free-fall time scale of the star, which reads\cite{2018pgrb.bookZ}
$t_{\rm ff}\sim (\frac{3\pi}{32\rm G\bar{\rho}})^{1/2}\sim 210~\rm s~(\frac{\bar{\rho}}{100~\rm g~\rm cm^{-3}})^{-1/2}$, where $\bar{\rho}$ is the mean density of the accreted matter. One can see that for the typical density $\bar{\rho}\sim 100~\rm g~\rm cm^{-3}$ of a massive star, the duration should be long. If the $\sim 1$ s duration of GRB 200826A is indeed the total duration of the central engine, one immediately poses a lower limit on the density of the accreted materials, i.e.
$ \bar\rho > 4.4\times 10^6 \ {\rm g \ cm^{-3}} \ T_{90}^{-2}$.
This would directly rule out any massive star progenitor with an accretion-powered engine. 

There are two ways to get around this argument. The first possibility is indeed to introduce a compact object rather than a massive star at the central engine. One category of model invokes a white dwarf (WD) engine. Since single WDs cannot make GRBs, one may consider various binary mergers invoking WDs\cite{Belc2002,Middleditch2004(1)}, e.g., WD-WD mergers, WD-NS mergers, and WD-BH mergers. One challenge of this scenario is whether such GRBs can still occur in star-forming regions in the host galaxies. Moreover, the detection of a possible supernova bump\cite{ahumada2020}
would disfavor this model. The other possible channel is that GRB 200826A's progenitor was a massive star, but it collapsed some time before the GRB was made. Such a scenario, known as ``supranova'', has been discussed in the literature\cite{vietri-stella98}. 
The GRB is produced by the implosion of a supermassive neutron star, probably triggered by the spindown or fallback accretion of the neutron star. As the GRB jet is launched, there are no more low-density materials surrounding the source, so that the duration is short. The source is still in the star forming region. One challenge of this model is to account for the relatively shallow decay in the X-ray afterglow, which seems to require additional energy injection during the afterglow phase (Methods). The possible supernova bump\cite{ahumada2020} also requires that the delay time from massive star core collapse to GRB jet launching cannot be much longer than a day. 

The second possibility is that the true duration of the central engine is long. However, during the majority of the active central engine time, the $\gamma$-ray emission is below the detection threshold of GRB detectors. Such a ``tip-of-iceberg'' interpretation may apply to other short-duration, low-$f$ Type II GRBs such as GRB 090426 (refs. \cite{levesque10,lv14}). However, the large $f$ value of GRB 200826A makes this simple interpretation unlikely. One may consider two modified scenarios. One is that the total central engine timescale $\Delta t_{\rm eng}$ is long. However, the majority of the time is used for the jet to penetrate through the stellar envelope in a timescale $\Delta t_{\rm jet} \sim 10$s. The observed GRB duration is $\Delta t_{\rm GRB} = \Delta t_{\rm eng} - \Delta t_{\rm jet}$, which could be as short as $\sim$ 1 s (ref. \cite{Bromberg2012}). Within this interpretation, one needs to introduce a coincidence between $\Delta t_{\rm eng}$ and $\Delta t_{\rm jet}$. This possibility has been emphasized also in ref. \cite{ahumada2020}. Alternatively, one may envisage that the engine indeed lasted a duration much longer than 1 s, but during the majority of the time the jet may carry heavy baryon loading and is mildly relativistic or non-relativistic so that no bright $\gamma$-rays could be produced during this period of time. One possibility is that the engine is a new-born magnetar which initially injects baryon-loaded, neutrino-driven wind\cite{metzger11}. The advantage of this model is that the X-ray plateau observed in this burst (Methods, \textcolor{black}{Extended Data Figure 6}) may be interpreted as the spindown of such a new-born magnetar. A short duration for this scenario may be achieved by invoking differential-rotation-induced magnetic bubbles as the mechanism of producing GRB prompt emission\cite{1998ApJ...505L.113K}. With a relatively high initial seed magnetic field strength $B \sim B_{14}\times 10^{14}$ G, the total duration may be estimated as $t\sim 1 B_{14}^{-1}\Omega_{4}^{-1} ~\rm s$, where $B_{14}=B/10^{14}$ G and $\Omega_4$ is the characteristic differential angular rotation speed in units of $10^4~\rm s^{-1}$ (ref. \cite{1998ApJ...505L.113K}).

\begin{table*}
\vspace{-1.5cm}

\centering
 \caption{Properties of GRB 200826A}
 \begin{tabular}{c|c}

 \hline \hline
 Duration [$T_{90}$] (10-800 keV) & $0.96_{-0.07}^{+0.06}~\rm s$ \\
 $f$-parameters [$f$] & $7.58\pm 1.23$\\
 Hardness Ratio (50-300/10-50 keV) & $0.803$ \\
 Spectral photon index [$\alpha$] & $ -0.68\pm 0.05$\\
 Spectral peak energy [$E_{\rm p}$] & $ 120.29_{-3.67}^{+3.93}$ $\rm keV$\\
 Isotropic energy [$E_{\rm \gamma,iso}$] & $ 7.09\pm 0.28\times10^{51}$ $\rm erg$\\
 Total fluence & $4.85\pm 0.19\times10^{-6}$ $\rm erg$ $\rm cm^{-2}$\\
 $\epsilon$ - parameter & $0.33$ \\
 Time lag(10-20 keV $\sim$ 250-300 keV) &$ 0.157{\pm 0.051}~\rm s$\\
 Redshift [$z$] & $0.7481\pm 0.0003$ \\
 Offset [$R_{\rm off}$] & $2.6$ $\rm kpc$ \\
 Half light radius [$R_{50}$] & $5.8$ $\rm kpc$ \\
 Normalized offset [$R_{\rm off}$/$R_{50}$] & $0.45$ \\
 Cumulatve light fraction [$F_{\rm light}$] & $0.79$ \\
 $\mathcal{O}$ calculated with(out)$f$ & {$-3.1$(1.5)} \\
 log $O$(II:I)$_{\rm prompt}$ calculated with(out) $f$ & {$-5.2$(-0.57)} \\
 log $O$(II:I)$_{\rm host}$ & {$2.5$} \\
 Peak flux [$F_{\rm p}$] & $9.11_{-1.17}^{+1.47}\times 10^{-6}$ $\rm erg$ $\rm cm^{-2} \rm s^{-1}$\\
 Peak luminosity [$L_{\rm \gamma,p,iso}$] & $1.41_{-0.21}^{+0.23}\times 10^{52}$ $\rm erg$ $\rm s ^{-1}$\\
 Star formation rate [SFR]& >$1.44\ \rm M_{\odot}~yr^{-1}$ \\
 Stellar mass [$M_*$] & $\rm 4.1 \pm 2.9 \times 10^9\ M_{\odot}$. \\
 Specific star formation rate [sSFR] & > $\rm 0.35\ Gyr^{-1}$\\
 Optical transient location & $\rm 00h 27m 08.5s\quad +34^{\circ} 01^{\prime} 38^{\prime \prime}.3$ \\
 SN-association & likely\\
 
 \hline \hline 
 \end{tabular}
 \label{tab:my_label}
 \vspace{0.5cm}
\end{table*}


\begin{addendum}
\item 
 B.B.Z acknowledges support by the National Key Research and Development Programs of China (2018YFA0404204), the National Natural Science Foundation of China (Grant Nos. 11833003, U2038105), and the Program for Innovative Talents, Entrepreneur in Jiangsu. Y.Z.M. is supported by the National Postdoctoral Program for Innovative Talents (grant No. BX20200164). This work was supported in part by the Natural Science Foundation of China (Grant Nos U1831135[XHZ], 11922301[HJL], 12041306[YL]), the Guangxi Science Foundation (2017GXNSFFA198008[HJL]), and the Bagui Young Scholars Program[HJL]. We also acknowledge the use of public data from the Fermi Science Support Center (FSSC).
 \item[Author Contributions] 
BBZ and HJL initiated the study. BBZ and BZ coordinated the scientific investigations of the event. BBZ, ZKL, ZKP, YL, HJL, JY, YSY, YHY, YZM and JHZ processed and analyzed the data. AJCT and YDH carried out the GTC optical observations. JRM, XHZ and JMB carried out the Lijiang 2.4m optical observations. XGW and EWL carried out the LCOGT observations. BZ and ZGD contributed to the theoretical interpretations to the event. BBZ and BZ wrote the paper with the contributions from all coauthors.

 \item[Competing Interests] The authors declare that they have no competing financial interests.

\item[Correspondence] Correspondence and requests for materials
should be addressed to B.-B. Zhang (bbzhang@nju.edu.cn) and B. Zhang (zhang@physics.unlv.edu)

\end{addendum}

\clearpage

\newpage

\section*{Methods}

\subsection{Fermi data analysis.}

We extracted photon events collected by the Sodium Iodide (NaI) scintillation detector nb whose boresight has the smallest angle relative to the burst location and conduct the following temporal and spectral analysis as described in ref.\cite{2011ApJ...730..141Z}.

\textit{Light curve analysis}: We subtracted the background from the light curves using the approach described in ref.\cite{2018NatAs...2...69Z}. From the sharp peak from $T_{0} + 0.09\rm \hspace{4pt} s$ to $T_{0} + 1.05 \rm \hspace{4pt} s$ (Figure \ref{fig1}a), we obtained its $T_{90}$ $\sim$ $0.96~\rm s$, representing the time between the epochs when 5$\%$ and 95$\%$ of the total fluence are collected.
\label{lightcurve}

\textit{Spectral analysis}: We conducted both the time-integrated and time-resolved spectral analysis during the $T_{90}$ time interval. The time-dependent spectral analysis is performed by slicing $T_{90}$ into 19 segments with a bin size of 0.05 seconds. Each segment is guaranteed to have at least 20 net photon counts. The background spectra were obtained by modeling the data in the signal-free time interval using the baseline method\cite{2018NatAs...2...69Z}. We used $GBM\hspace{3pt} Response\hspace{3pt} Generator$ (obtained from {\url{https://fermi.gsfc.nasa.gov/ssc/data/analysis/rmfit/gbmrsp-2.0.10.tar.bz2}}) to generate the detector response matrices (DRMs). Finally, we adopted $McSpecFit$\cite{2018NatAs...2...69Z} to fit both preprocessed spectra and plotted the best fitting results. The mainstream spectra fitting models, such as the single power law (PL), cutoff power law (CPL), band function (Band), black body (BB), and their hybrids were utilized to fit both spectra. By assessing the fitting performance via the combination of Profile-Gaussian likelihood (PGSTAT/dof) and Bayesian information criterion (BIC)\cite{2018NatCo...9..447Z}, the CPL model was favored for best representing the properties of both time-integrated and time-resolved spectra. The CPL model is described in the form of 
\begin{equation}
 N(E)=A\Big(\frac{E}{100~\rm keV}\Big)^{\alpha}{\rm exp}\Big[\frac{-(\alpha+2)E}{E_{\rm p}}\Big],
 \label{N(E)}
\end{equation}
 where $\alpha$ is the photon spectral index, $A$ is the normalized coefficient, and $E_{\rm p}$ is the peak energy of the spectrum. We fitted time-dependent spectra in 19 aforementioned time intervals using the CPL model. Fitting results demonstrate an intensity tracking pattern of spectral evolution, which is shown in \textcolor{black}{Extended Data Figure 3}. Time-integrated fitting results are exhibited in \textcolor{black}{Extended Data Figure 2}. We list the results in \textcolor{black}{Extended Data Figure 1}.

\subsection{Amplitude parameter.} The $f$ parameter is defined by $f \equiv \frac{F_{\rm p}}{F_{\rm b}}$, where $F_{\rm p}$ and $F_{\rm b}$ correspond to the peak flux and average background flux of a GRB, respectively. In order to distinguish intrinsically short GRBs from disguised short GRBs due to the tip-of-iceberg effect, an effective amplitude parameter $f_{\rm eff}=\frac{F_{\rm p'}}{F_{\rm b}}$ for long GRBs is defined, which is the $f$ value of ``pseudo GRBs'' whose flux is re-scaled from the initial flux so that $T_{90}$ above the background flux is shorter than 2 s(ref.\cite{lv14}) . By comparing the $f - T_{90}$ diagram of short GRBs and the $f_{\rm eff} - T_{90}$ diagram of long GRBs, one can see that most of intrinsically short GRBs possess relatively large $f$ than $f_{\rm eff}$ of long GRBs. A disguised short GRB would have $f < 1.5$. GRB 200826A has { $f=7.58 \pm 1.23$,} which is much greater than the range for a disguised GRB. As a result, this GRB is a genuinely short GRB.

\subsection{Hardness ratio.} Long GRBs are usually softer than short GRBs, suggesting that they have a lower hardness ratio (HR). We define HR as the photon counts ratio between 50-300 keV and 10-50 keV energy bands. The GRB sample (from ref.\cite{2017ApJ...848L..14G}) and GRB 200826A are marked in Figure \ref{fig1}c by points and star, respectively.

\subsection{Amati relation.} 
The correlation between the GRB isotropic energy $E_{\rm\gamma,iso}$ and the rest-frame peak energy $E_{\rm p,z}=(1+z)E_{\rm p}$ was discovered by Amati et al.\cite{2002A&A...390...81A}, and can be written as ${\rm log~}E_{\rm p,z}=a+b{\rm log~}E_{\rm \gamma,iso}$. Typically, long and short GRBs follow different tracks in the $E_{\rm p,z}-E_{\rm\gamma,iso}$ diagram. We over-plotted the $E_{\rm p,z}-E_{\rm\gamma,iso}$ diagram (Figure \ref{fig2}a) using GRB samples with known redshift from refs. \cite{2009ApJ...703.1696Z1} and \cite{2002A&A...390...81A}. Utilising the MCMC method, the optimal fitting parameters and $3\sigma$ uncertainties are constrained. We obtain $a = -25.01^{+3.91}_{-4.25}$ and $b = 0.52^{+0.08}_{-0.07}$ for long GRBs, and $a = -19.92^{+9.18}_{-7.60}$ and $b = 0.45^{+0.15}_{-0.18}$ for short GRBs. The luminosity distance is calculated with a cosmological parameter set of $H_0=67.7$ km s$^{-1}$ Mpc$^{-1}$, $\Omega_{\rm m}=0.307$ and $\Omega_{\rm \Lambda}=0.693$ throughout this paper.

\subsection{\bf$\epsilon$ parameter.} With the parameter {\bf$\epsilon$} = $E_{\rm \gamma,iso,52}/E_{\rm p,z,2}^{5/3}$, where $E_{\rm \gamma,iso,52}=E_{\rm \gamma,iso}/10^{52}~\rm erg$ and $E_{\rm p,z,2}=E_{\rm p,z}/100~\rm keV$, a classification method \cite{2010ApJ...725.1965L} was proposed to clearly divide GRBs with high $\epsilon$ ($\epsilon > 0.03$) and low $\epsilon$ ($\epsilon < 0.03$) at the separation line $\epsilon$ $\sim$ 0.03. The two classes nicely match Type II and Type I GRBs, respectively. We computed $\epsilon$ of GRB 200826A and plotted it on the ${\rm log ~}T _{\rm 90,z}$ - ${\rm log ~}\epsilon$ panel (Figure \ref{fig2}b). The classification of a GRB can be quantitatively evaluated by a GRB distribution probability, which is defined as the number of GRBs falling inside an area of $d{\rm log~}\epsilon$ by $d{\rm log~} T_{90,z}$ centered at ${\rm log~}\epsilon$, ${\rm log~} T_{90,z}$ divided by the total number of the GRB samples. We assume this probability of each GRB in our sample $p_i$ has a normalized log-normal probability distribution $p$(${\rm log~} \epsilon, \delta {\rm log~} \epsilon $). We calculated the integral of $p_{i}$ over the range [${\rm log~}\epsilon - {d\rm log~} \epsilon/2, {\rm log~}\epsilon + {d\rm log~} \epsilon/2$] and summed over the entire GRB sample to obtain the probability $P({\rm log~}\epsilon, {\rm log~}T_{\rm 90,z})$. Dark grey ($P({\rm log~}\epsilon, {\rm log~}T_{\rm 90,z})>0.1$) and light grey ($P({\rm log~}\epsilon, {\rm log~}T_{\rm 90,z})>0.003$ at 3$\sigma$ confidence level) regions in Figure \ref{fig2}b mark probability contours of GRB clustering, respectively. GRB 200826A falls into the high $\epsilon$ region with $P({\rm log~}\epsilon, {\rm log~}T_{\rm 90,z})>0.003$, which suggests that it is of a Type II origin.

\subsection{Spectral lag.} For most GRBs, time lags among different bands, also known as spectral lags, exist in the light curves\cite{2017ApJ...834L..13W} due to the fact that photons with different energies arrive the observer at different times. We used the cross-correlation function (CCF)\cite{2000ApJ...534..248N,2010ApJ...711.1073U} to calculate the time lags of light curves among different energy bands between $T_{0}$ + $0.5 \rm \hspace{4pt}s$ and $T_{0}$ + $ 1.6 \rm \hspace{4pt}s$ following the method described in ref.\cite {2012ApJ...748..132Z}. We obtain non-zero 
spectral lags for GRB 200826A, which are shown in \textcolor{black}{Extended Data Figure 4}.

{\subsection{X-ray Observations.} 
The Swift/XRT has performed a follow-up observation for a duration of $\sim6\times 10^{4}~\rm s$ after the GBM trigger. The observation (obtained from {\url{https://www.swift.ac.uk/xrt_live_cat/00021028/}}) provided an intrinsic column density of $6.0\times 10^{20}$ cm$^{-2}$. The X-ray light curve is shown in the Extended Data Figure \ref{fig:afterglow}. We fitted the light curve using a broken power-law model and obtained the decay slopes as $\sim -1.5$ and $\sim -0.4$, and a break at $\sim 10^5$ s. }

\subsection{Optical transient.} 

An optical counterpart of GRB 200826A was located by Zwicky Transient Facility
(ZTF) and followed up by several ground and space facilities. We summarized our own observations performed by the Gran Telescopio CANARIAS (GTC), Lijiang 2.4-meter telescope, and LCOGT network observatory in Extended Data Table 2. Other sets of optical observations can be found in ref.\cite{ahumada2020}

\subsection{Host galaxy observations and properties.} 

We observed the host galaxy with the Gran Telescopio CANARIAS (GTC) in $r$-band, LCOGT network observatory in $I$-band. Those results allow us to perform the following analyses on the host galaxy properties. 
\begin{itemize}
 \item $R_{\rm 50}$. The ZTF position is used to estimate the half-light radius $R_{\rm 50}$. We used the PyRAF/$ellipse$ routine to generate elliptical intensity isophotes, constructed the total flux profile as a function of semi-major axis, and calculated the half-light radius, $R_{\rm 50}$, through interpolation. We obtained $R_{\rm 50}$ $\sim$ 0.77", corresponding to 5.7 kpc at redshift 0.7481. With a 0.35" offset mentioned by Ahumada, T. et al.,
 the normalized offset $R_{\rm off}/R_{50}=0.45$ is a typical value for LGRBs. 
 \item $F_{\rm light}$. We estimated the cumulative light fraction $F_{\rm light}$ with our GTC image. In the intensity versus total flux profile, we interpolated the mean intensity at the position of GRB 200826A to estimate the total flux of the fainter regions and $F_{\rm light}=F_{\rm fainter}/F_{\rm total}=0.79$. The offsets and $F_{\rm light}$ of GRB 200826A are typical for LGRBs.
 Comparatively, GRB 090426A has a smaller offset and larger $F_{\rm light}$ than GRB 200826A.
 \item SFR. We calculated the star formation rate (SFR) with near-UV (NUV) observations (\textit{and emission lines}). 
 Since the rest frame wavelength of the $g$-band is in NUV, we are able to approximate the SFR with the Galactic extinction corrected $g$-band magnitude from DESI, as {8.7} $M_{\odot}$ yr$^{-1}$ based on Kennicutt 1998\cite{kennicutt1998}. 
 The [OII]$\lambda$3727 emission line flux present in the GTC spectrum\cite{ahumada2020} is $3.81 \times 10^{-17}$ erg s$^{-1}$ cm$^{-2}$, which corresponds to a luminosity {$1.03\times10^{41}$} erg s$^{-1}$. The SFR estimated with ref. \cite{kennicutt1998} using the [OII] emission line is $(1.4\pm0.4)\times 10^{-41} L_{\rm [OII]}> {1.44}\ M_{\odot}$ yr$^{-1}$. Emission lines represent
 star formation rate with age 0-10 Myr, which is more related to the origin of LGRBs, hence serves as a better indicator of distinguishing LGRBs and SGRBs. We thus used the SFR estimated with [OII] emission line to obtain the Odds O(II:I) (see below).
 \item Stellar mass. To estimate the stellar mass, we used the Code Investigating GALaxy Emission (CIGALE) \cite{2009A&A...507.1793N} to perform Spectral Energy Distribution (SED) fitting in $g$, $r$, $i$ and $z$ bands after considering the Galactic extinction \cite{2011ApJ...737..103S}. Several models, including the {\it sfhdelayed} star formation history model with an initial SFR of 0.1 $M_{\odot}$ yr$^{-1}$, {\it bc03} stellar population model\cite{2003MNRAS.344.1000B} with a Chabrier initial mass function, {\it dustatt\_calzleit} dust attenuation model with the UV bump centroid to be 217.5 nm, {\it dale2014} dust emission model\cite{2014ApJ...784...83D} with various AGN fractions, were taken into account. Our results yield a stellar mass of $M_*= (4.1 \pm 2.9) \times 10^9 M_{\odot}$. Correspondingly, the specific star formation rate (sSFR) can be calculated as $> {0.35}$ Gyr$^{-1}$.

\end{itemize}

\subsection{$\mathcal{O}$.} A naive Bayesian scheme has been proposed by Li et al., 2020\cite{liye2020} to determine the physical categories of GRBs based on multi-wavelength properties. The parameter $\mathcal{O}={\rm log\ O(II:I)}+0.7$ is the logarithmic Type II to Type I probability ratio corrected for the potential correlation between parameters.
We calculate $\mathcal{O}$ based on the parameter sets from the prompt emission and host galaxy properties. A GRB is classified as a Type II or Type I if $\mathcal{O}$ is positive or negative. Without inclusion of the $f$ parameter, we obtain a positive value $\mathcal{O}$ = 1.5, placing GRB 200826A in the Type II rather than Type I GRB category. With the method proposed in ref.\cite{liye2020}, we further separated the parameter sets into groups of prompt emission and host galaxy to compute log O(II:I)$_{\rm prompt}$ and log O(II:I)$_{\rm host}$, respectively, and place GRB 200826A on the distribution diagram (Figure \ref{fig3}f).

\subsection{Data Availability}
 Processed data are presented in the tables and figures in the paper. Source and optical observational data are available upon reasonable requests to the corresponding authors. The {\it Fermi}/GBM data are publicly available at \url{https://heasarc.gsfc.nasa.gov/FTP/fermi/data/}.
 
\subsection{Code Availability}
Upon reasonable request, the code (mostly in python) used to produce the results and figures will be provided.

\clearpage
\setcounter{figure}{0}
\setcounter{table}{0}

\captionsetup[figure]{name={\bf Extended Data Figure}}

\begin{figure}
\vspace{-4cm}
\hspace{-3cm}
\includegraphics[width=8.0in]{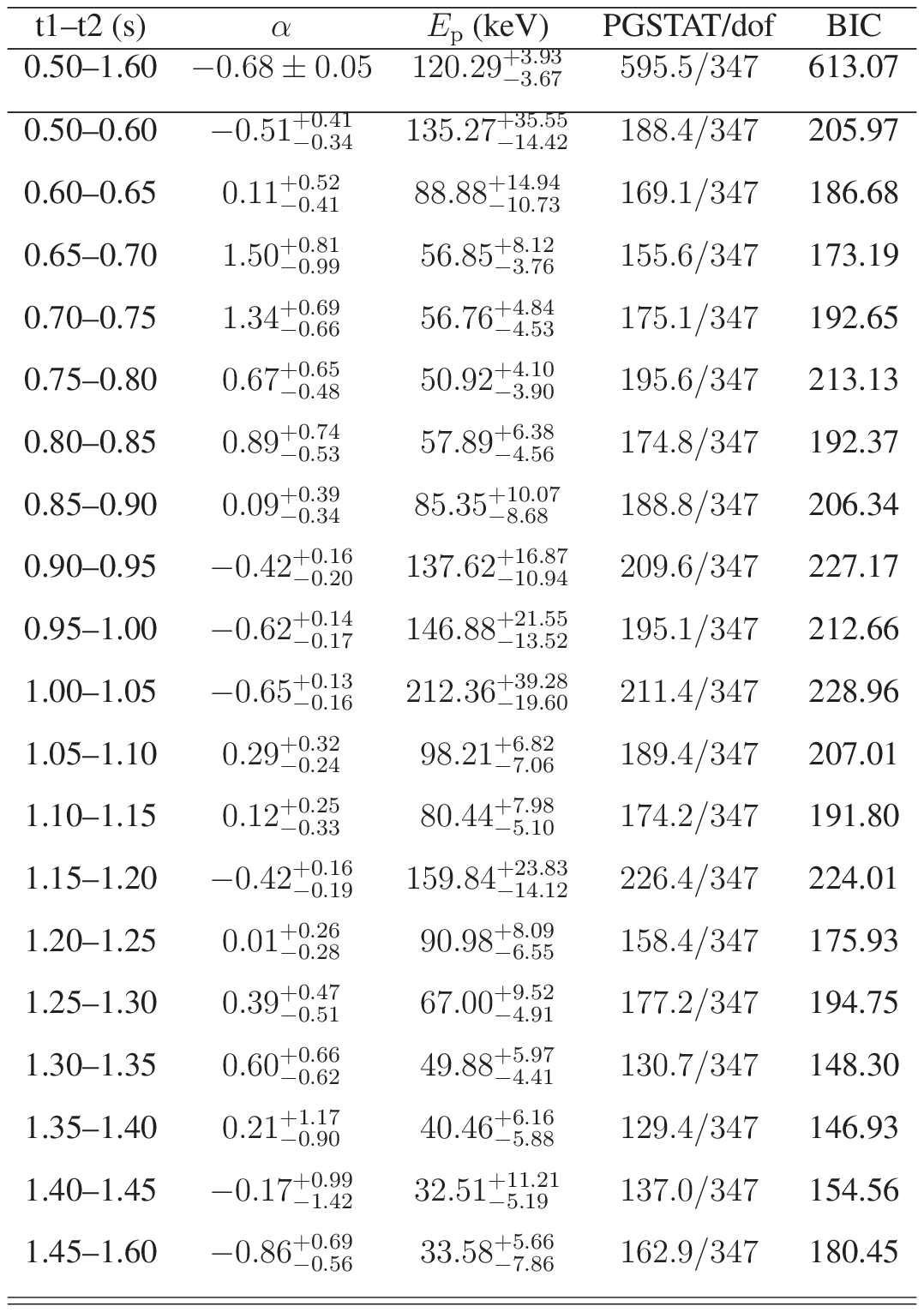}
\vspace{-8cm}
\caption{Time-integrated and time-resolved spectral fitting results of GRB 200826A with \textit{cutoff power law} model} 
\label{fig:spec1}
\end{figure}
\clearpage

\begin{figure*}
\vspace{-2.5cm}
\includegraphics[width=1.\linewidth]{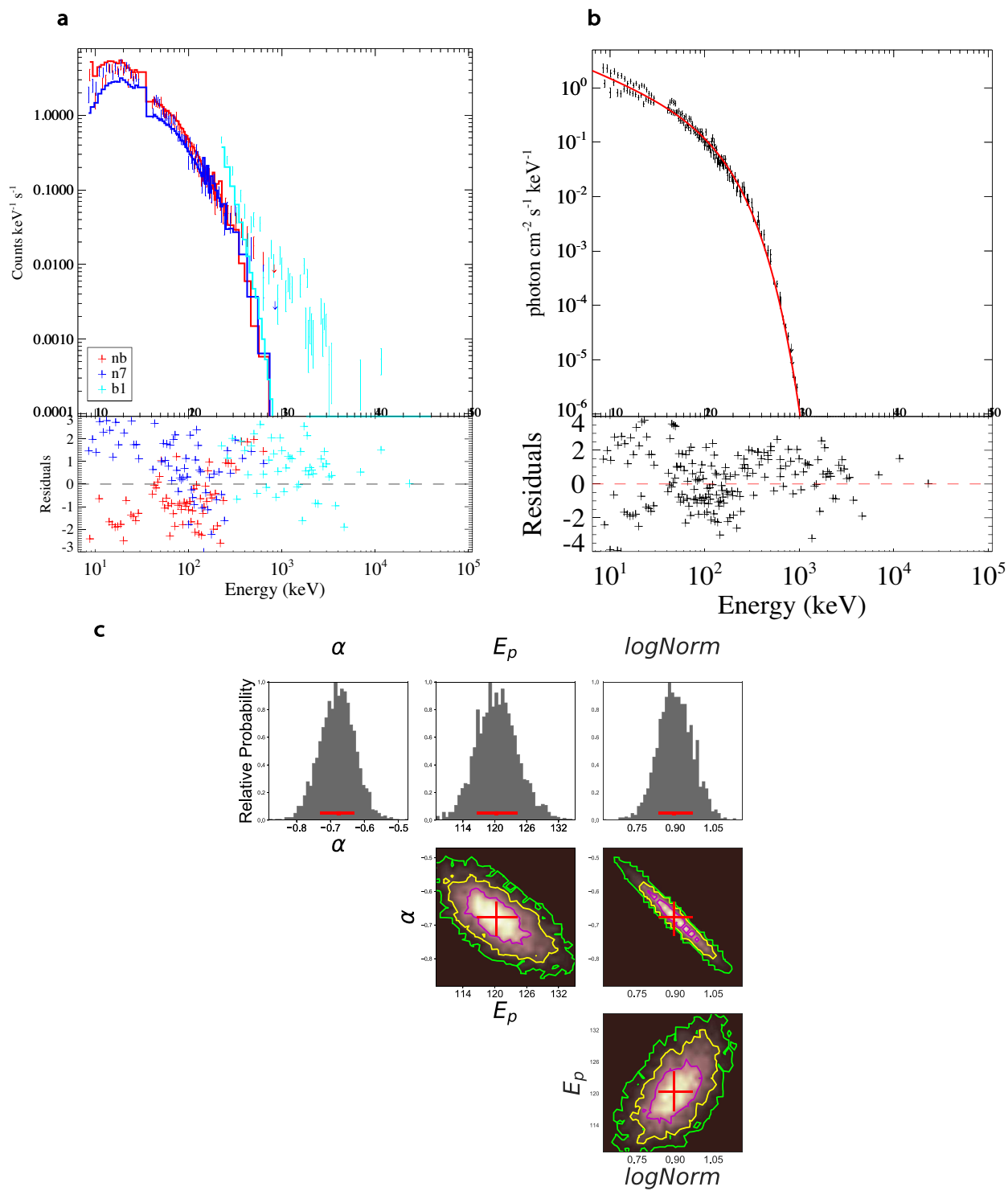}

\caption{Spectral fitting results within the interval of $T_{0}+0.50~\rm s$ $\sim$ $T_{0}+1.60~\rm s$. {\bf a}, observed photon count rate spectra and fitted model of nb(red), n7(blue) and b1(sky blue) detectors. {\bf b,} de-convolved spectra(black points) and best-fit power law model(red line). {\bf c,} corner plots and histograms show one and two-dimensional posterior probability distributions of cutoff power-law model parameters at 1-$\sigma$ (purple contours), 2-$\sigma$(yellow contours) and 3-$\sigma$(green contours) confidence levels. Red error bars and crosses represent best-fit values with 1-$\sigma$ uncertainties.}
\vspace{-2cm}
\label{fig:corner}
\end{figure*}
\clearpage


\begin{figure}
\vspace{-2cm}
\hspace{-2cm}
\includegraphics[width=8.0in]{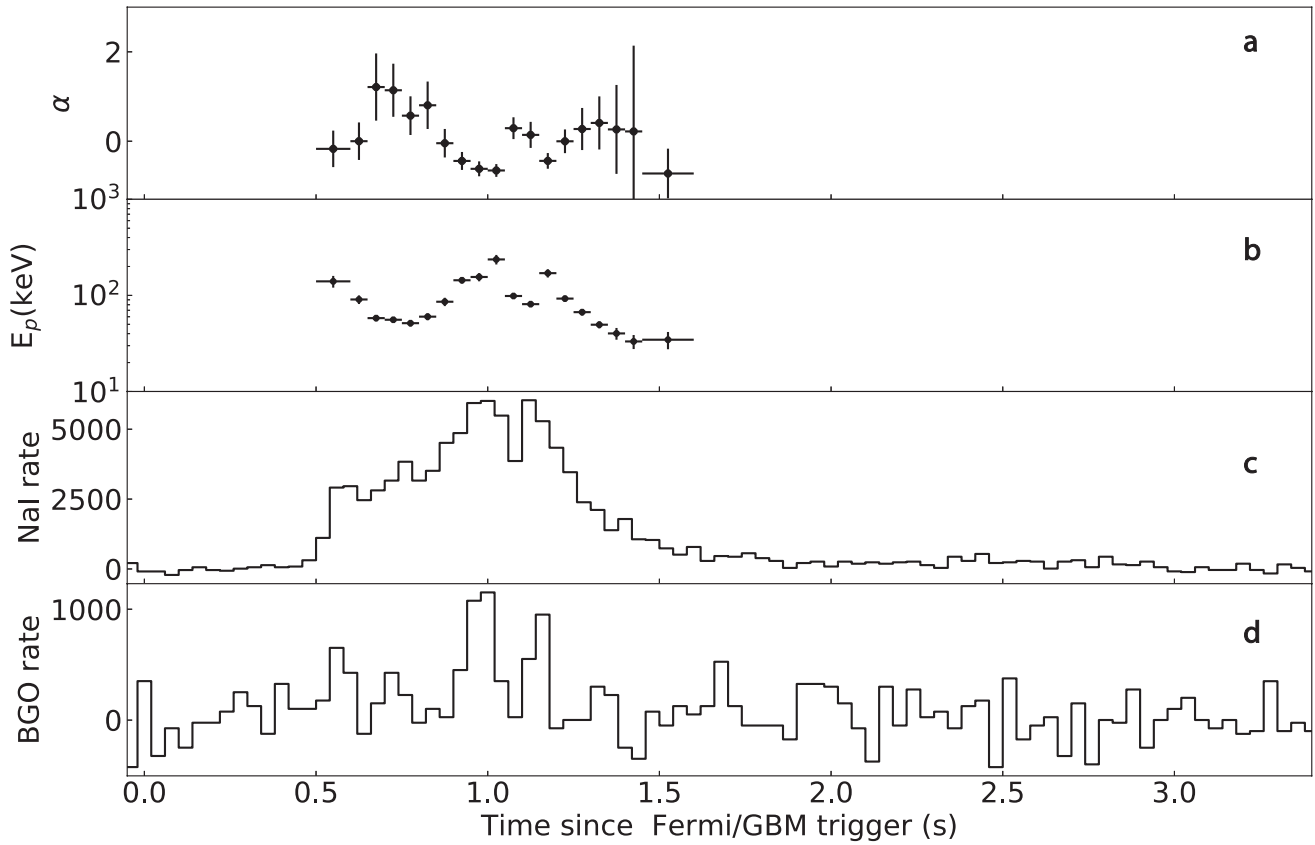}
\caption{Spectral evolution of GRB 200826A. Panels a and b show evolution of the photon index ($\alpha$) and spectral peak energy ($E_{\rm p}$) of the cut-off powelaw model, respectively. Panels c and d show the NaI and BGO light curves. All error bars represent 1-$\sigma$ uncertainties.} 
\label{fig:spec}
\end{figure}

\begin{figure*}

\includegraphics[width=1.1\linewidth]{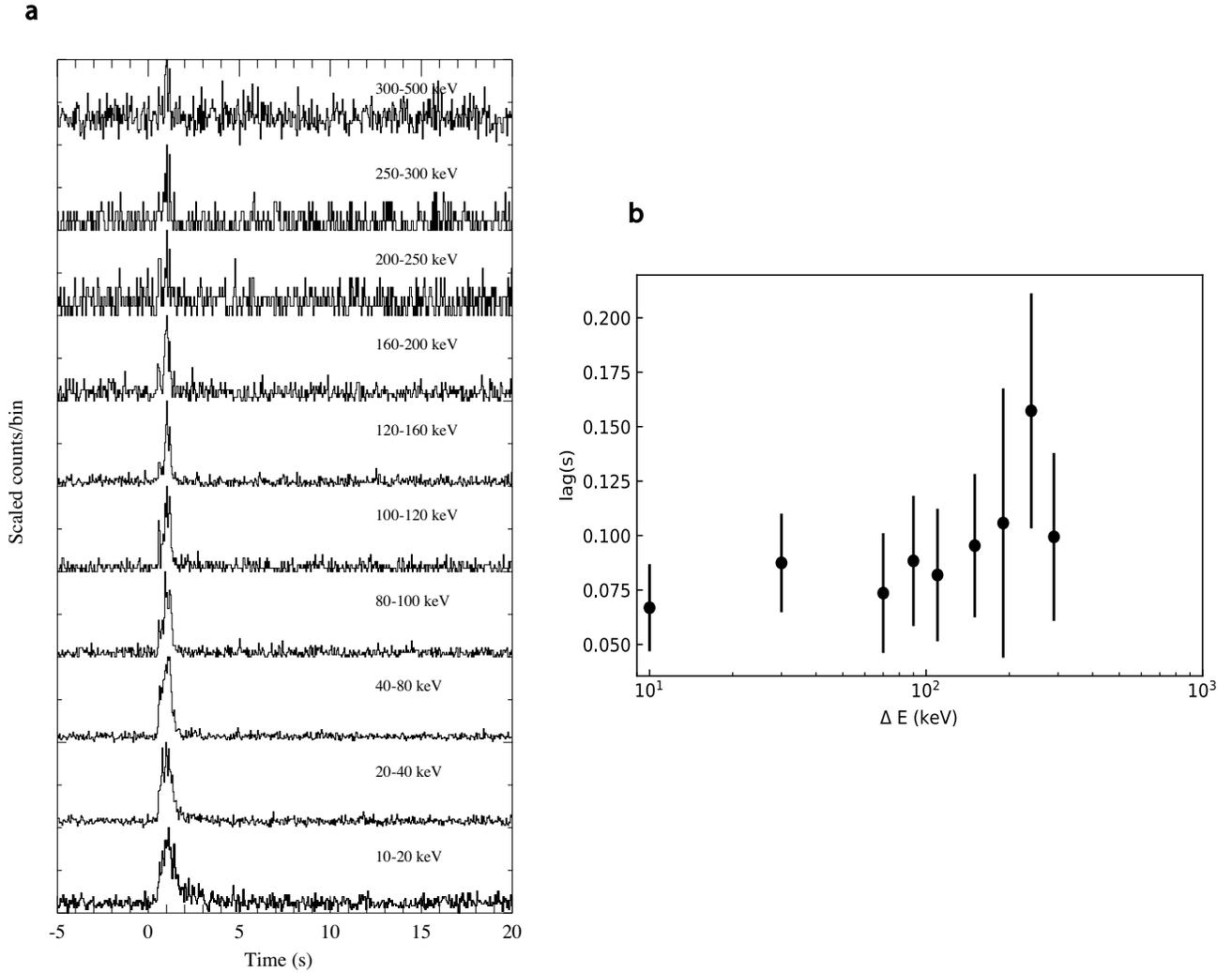}
\caption{Spectral lag calculation. {\bf a,} light curves in different energy bands(10-20 keV $\sim$ 300-500 keV) which are used to calculate lags. {\bf b,} energy dependent spectral lag between the lowest energy(10-20 keV) band and any higher energy band. All error bars represent 1-$\sigma$ uncertainties.}
\vspace{-1cm}
\label{fig:lags}
\end{figure*}
\begin{figure*}
\hspace{-7.0cm}

\includegraphics[width=1.0\linewidth]{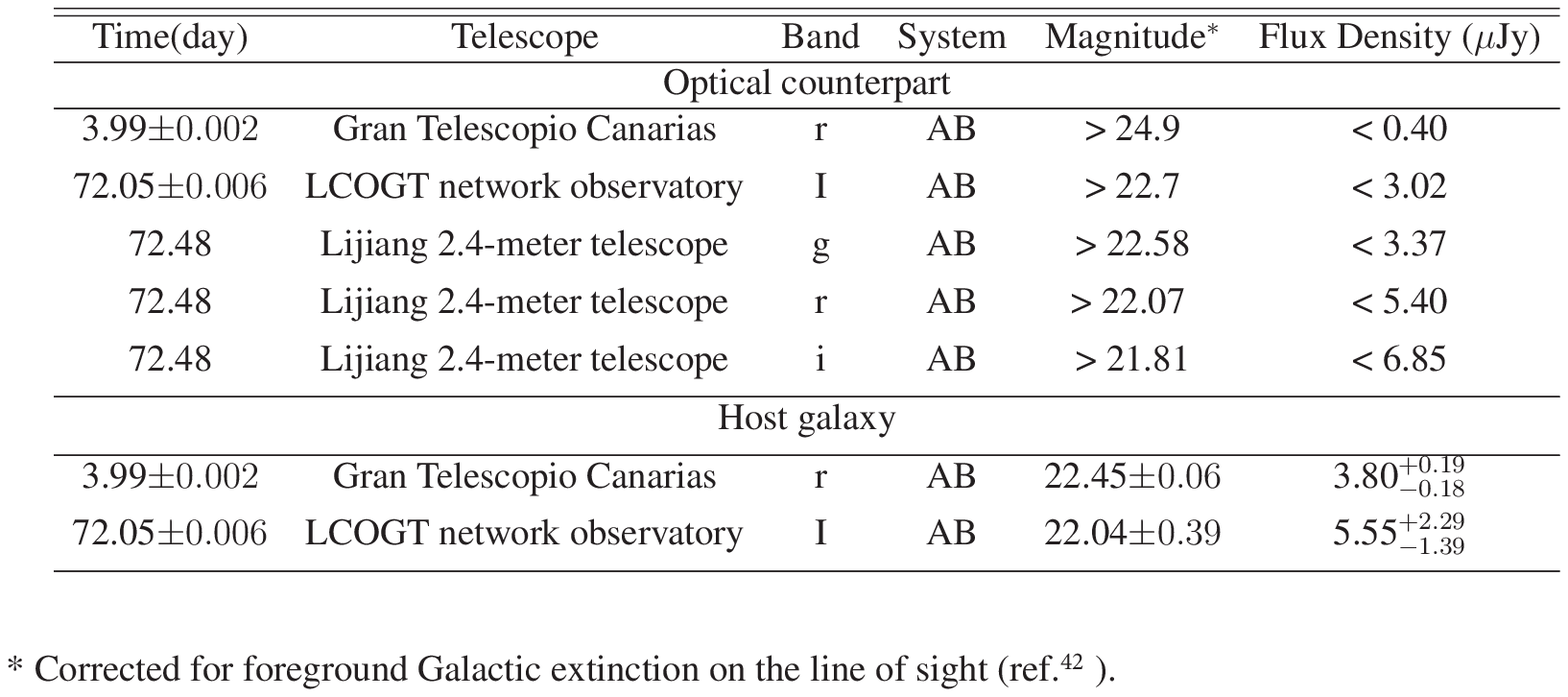}
\vspace{-4.5cm}
\caption{Observations of optical counterpart and host galaxy of GRB 200826A.}
\vspace{-1cm}
\label{fig:lags1}
\end{figure*}

\begin{figure}
\hspace{-1cm}
\vspace{3 cm}
\centering
\includegraphics[width=6.5in]{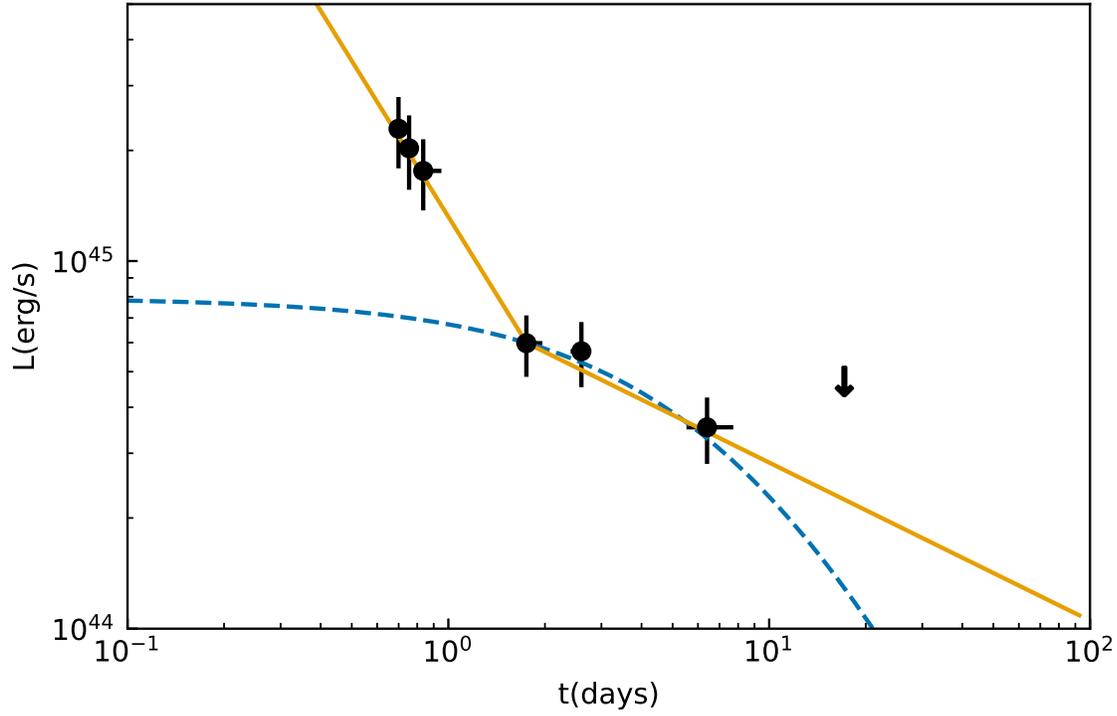}
\caption{X-ray afterglow of GRB 200826A. Black points represents the observed X-ray flux in 0.3-10 keV. The solid orange line shows the broken power-law fitting with slopes $\alpha$ = -1.41$_{-0.12}^{+0.24}$, $\beta$= -0.43$_{-0.22}^{+0.17}$ and a break at $t_{\rm b}$ = $1.51^{+0.31}_{-0.30} \times 10^{5}$ s. The dashed blue line represents the magnetar spin-down energy injection model parameterized by $L_{\rm sd} = L_{0}(1+t/\tau)^{-2}$ with $L_{0} = 10^{44.9}~\rm erg~s^{-1}$ and $\tau = 1.0\times 10^{6}~\rm s$. All error bars represent 1-$\sigma$ uncertainties. The upper limit is at the 3-$\sigma$ level. } 
\label{fig:afterglow}
\end{figure}

\end{document}